\documentclass[superscriptaddress,twocolumn,aps,prl]{revtex4-1}

\usepackage{amsmath}
\usepackage{amstext}
\usepackage{graphicx}  
\usepackage{array}

\begin{document}

\title{Probing For Non-Gravitational Galactic-Range Dark Matter Interactions}
\author{M.P. Ross}
\author{ S.K. Apple}
\author{ E.A. Shaw}
\author{ C. Gettings}
\author{ I.A. Paulson}
\author{ J.H. Gundlach}
\affiliation{Center for Experimental Nuclear Physics and Astrophysics, University of Washington, Seattle, Washington 98195, USA}

\begin{abstract}
Dark matter remains a mystery in fundamental physics. The only evidence for dark matter’s
existence is from gravitational interactions. We operated a precision torsion balance experiment to search for non-gravitational, galactic-range interactions between ordinary matter in our lab and the Milky Way's dark matter. We find no evidence of such interaction and set upper bounds on its strength that are a fourfold improvement over previous limits.  These results constrain a variety of dark matter theories and test the equivalence principle towards dark matter.

\end{abstract}

\maketitle

\section{Introduction}
Dark matter has remained a persistent mystery in fundamental physics and cosmology. Decades of cosmological observations -- including galactic rotation curves \cite{rotcurve}, large-scale structure surveys \cite{des, desi}, and measurements of the cosmic microwave background \cite{planck}  -- have established that dark matter constitutes approximately 26\% of the universe's mass-energy content, far greater than the ~5\% contribution from ordinary matter \cite{planck}. The luminous matter of the Milky Way galaxy is embedded in a halo of dark matter that makes up most of the galaxy's mass~\cite{HUNT2025101721}. Despite this apparent abundance, the true identity of dark matter remains elusive and it is still unknown whether dark matter interacts in any other way beyond gravity.

\begin{figure}[!h]
    \centering
    \includegraphics[width=0.4\textwidth]{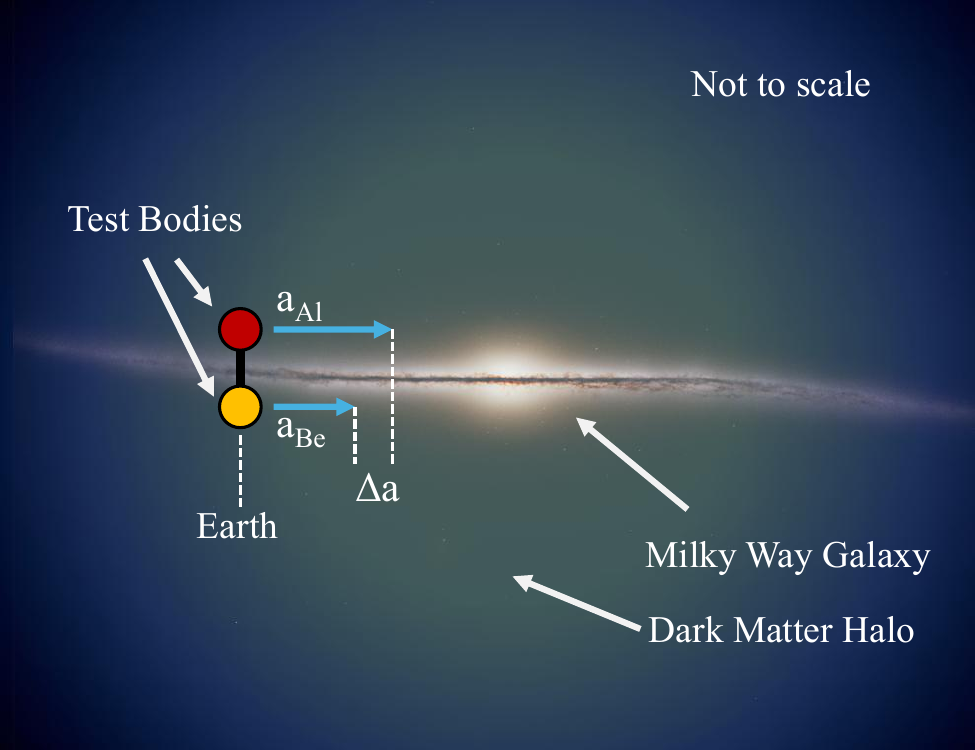}
    \caption{A schematic of the experimental setup in the galactic frame. The experiment searches for differential accelerations of two different materials towards the dark matter in the Milky Way galaxy. Adapted from ESA/Gaia/DPAC, Stefan Payne-Wardenaar \cite{milkyway}.}
    \label{cartoon}
\end{figure}

The prevailing paradigm envisions dark matter as a new class of non-baryonic, weakly-interacting particles \cite{bertone2010particle, arcadi2018waning}. However, every effort at detecting and producing particle dark matter has come up empty \cite{BOZORGNIA2024}. 

\begin{figure}[!h]
    \centering
    \includegraphics[width=0.5\textwidth]{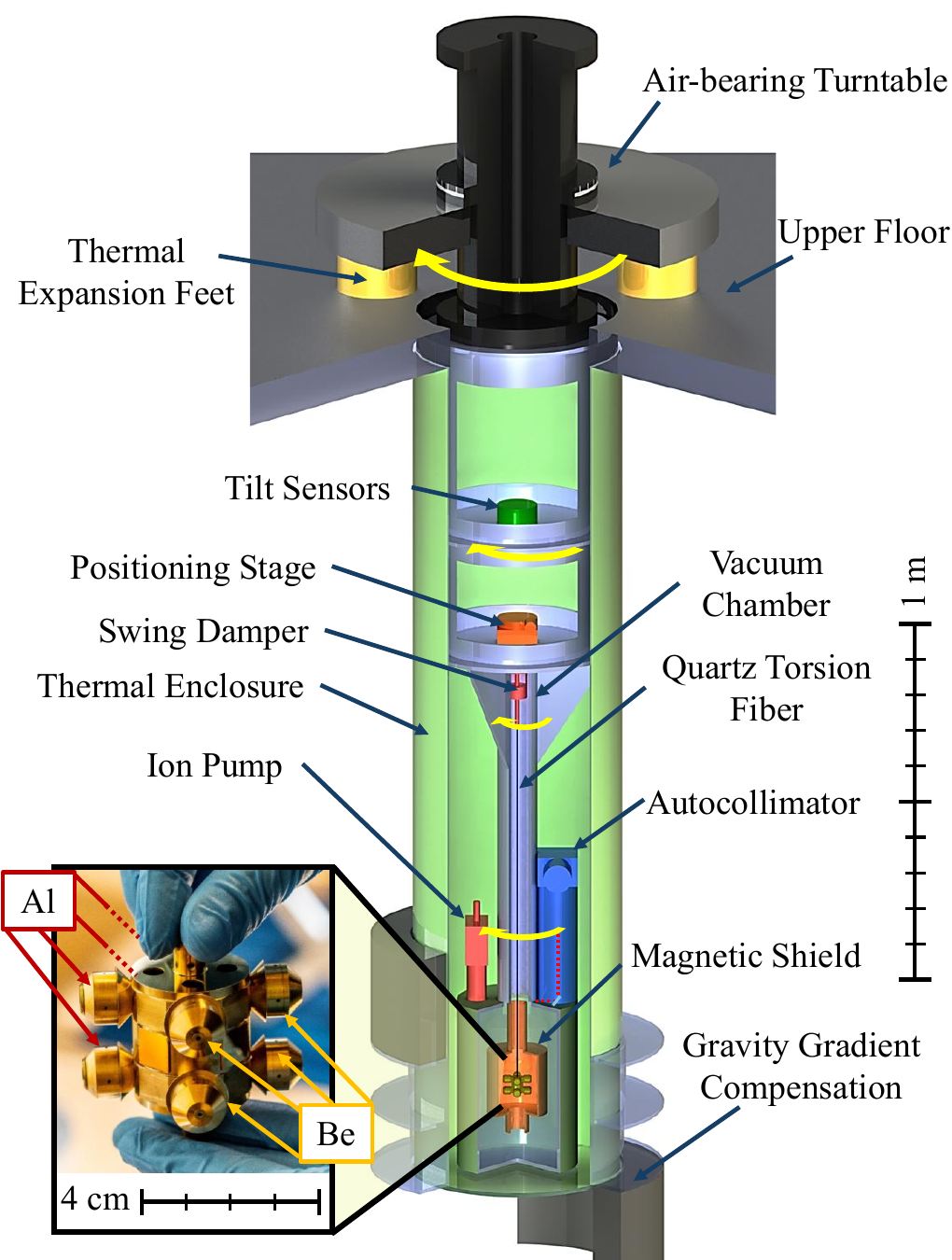}
    \caption{Schematic of the rotating torsion balance apparatus with embedded picture of the pendulum held before installation. The thermal shields, vacuum chamber, and magnetic shields have been cut away to show the torsion balance held inside. The torsion pendulum has four test bodies made of aluminum and four of beryllium in a dipole orientation. The pendulum is suspended by a quartz torsion fiber attached to a magnetic swing damper whose position can be actuated by the positioning stage. The vacuum chamber, torsion balance, autocollimator, and magnetic shield are suspended from the air-bearing turntable. The rotating tilt sensors are used to level the turntable by actuating the thermal expansion feet. The angle of the pendulum relative to the vacuum chamber is measured by the autocollimator. 
    }
\label{apparatus}
\end{figure} 

This invites serious consideration of nontraditional possibilities including: primordial black holes \cite{PBH}, sterile neutrinos \cite{sterile}, and entirely undiscovered phenomena.

We operated a laboratory experiment to search for non-gravitational, galactic-range interactions between ordinary matter and the Milky Way's dark matter. This experiment measured the acceleration difference between two different materials directed towards the center of the Milky Way galaxy, as shown in Figure~\ref{cartoon}. Resolving such minuscule differential accelerations required development of an ultra-sensitive torsion balance apparatus. To achieve this, we upgraded our previously constructed rotating torsion balance \cite{wagner2012torsion} with a fused quartz torsion fiber as well as a high-speed data acquisition and control system \cite{shaw2023equivalence}.

\section{Apparatus} 

Torsion balances are highly sensitive mechanical systems that have long been used for fundamental physics experiments~\cite{ADELBERGER2009102}. They are formed by a pendulum suspended from a thin fiber that acts as a weak torsional spring.

Our torsion pendulum consisted of a light-weight frame that held four test bodies made of beryllium and four made of aluminum. This formed a composition dipole oriented perpendicular to the torsion fiber (Figure \ref{apparatus}). The pendulum was designed with a high degree of symmetry to minimize coupling to gravitational gradients and other known interactions, such as electrostatics. The external surfaces were coated with gold to reduce surface charges. The pendulum was suspended from a 22-{\textmu}m thick, 1-m long fused quartz fiber \cite{shaw2022torsion}, which gave the balance a torsional period of 1472 seconds and a ring down time of $\sim$2 years.

The torsion balance was held inside a vacuum chamber maintained at  $< 0.2~\text{mPa}$ by an ion pump. At low frequencies ($< 1~\text{mHz}$), internal thermal noise \cite{thermal} of the torsion fiber limited the performance of the balance. The angle of the pendulum relative to the vacuum chamber was measured with an autocollimator. The autocollimator reflected a laser beam off one of the four mirrors attached to the pendulum. Table \ref{appTable} displays the relevant parameters of the torsion balance apparatus.

To avoid slowly varying noise (i.e. 1/f-noise), we modulated the signal of interest by suspending the entire torsion balance apparatus from a high-precision air-bearing turntable that was set to continuously rotate at 2.85~mrad/s  (2/3 of the torsional resonance frequency). The turntable was located on the upper floor of our laboratory with the rest of the apparatus extending to the floor below. The turntable's angle encoder was locked to an external clock which maintained a constant rotation rate by varying the torque applied with an eddy current motor \cite{shaw2023equivalence}. 

\begin{widetext}

\begin{figure}[!h]
\centering \includegraphics[width=\textwidth]{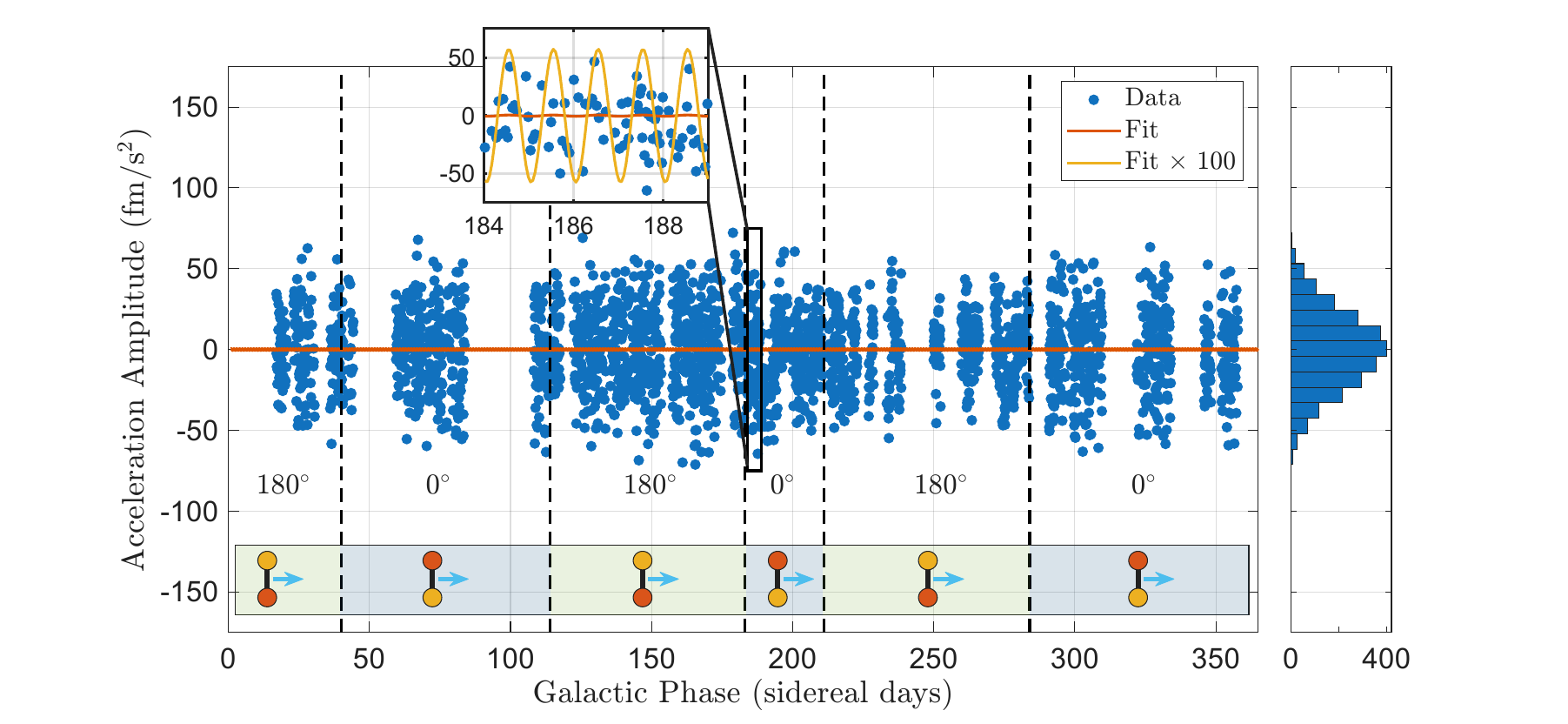}
\caption{Once per turntable revolution differential acceleration amplitudes along with the corresponding fit to the galactic basis functions. The mean of each data set has been subtracted. Gaps in the data were due to local construction activity. Data for sidereal days 189 to 285 was taken in 2024, while the remaining data was taken in 2025. The right panel shows the histogram of the differential acceleration amplitudes, the bottom panel shows the orientations of the pendulum with respect to the vacuum chamber over the data run. The zoom shows a five-day long segment and it's corresponding fit to the galactic basis functions.}
\label{time} 
\end{figure}

\end{widetext}

Misalignment of the rotation axis of the turntable and the torsion fiber axis (i.e. local vertical) can cause a parasitic once per turntable-revolution modulation by varying the torsion fiber attachment, translating the autocollimator beam, and/or changing the electrostatic environment around the pendulum. Such a modulation would not be correlated with the dark matter signal but could be a significant noise source. To mitigate this, we devised a control system that processed the signal from a pair of co-rotating tilt sensors to control the temperature, and thereby thermal expansion, of two of the three feet on which the turntable rests. Thermal expansion of the feet aligned the turntable rotation axis to local vertical \cite{shaw2023equivalence}.

The pendulum was built from strictly non-magnetic materials and machined with ceramic tools to avoid ferromagnetic impurities. Additionally, the pendulum was surrounded by three layers of co-rotating magnetic shielding and one non-rotating shield to minimize the effects of external magnetic fields. The inner-most magnetic shield was carefully demagnetized to avoid eddy current damping on the pendulum which preserved the high quality factor and low thermal noise.

To reduce effects of temperature variations, the vacuum chamber was surrounded by a co-rotating sheet metal shield, a concentric lab-fixed shield made of aluminum and insulating foam, and an enclosure made of polystyrene foam insulation. The torsion fiber was also surrounded by a thermally-isolated thick-walled copper tube. Finally, the experiment was located in an underground laboratory which was temperature controlled to $\pm 25~\text{mK}$ daily variation with a non-cycling air conditioner. 

\begin{table}
\centering
\begin{tabular}{|c|c||c|}
 \hline
 \multicolumn{3}{|c|}{Apparatus} \\
 \hline
 Parameter &  & Value \\
 \hline
 Fiber Diameter & \centering{$d$}  &\ $  22 $ {\textmu}m\\ 
 Fiber Length & \centering{$l$}  &\ $  0.94 $ m\\  
 Spring Constant & \centering{$\kappa$}  &\ $ 7.1 \times 10^{-10}$ N m/rad\\
 Quality Factor & \centering{$Q$}  &\ $ >1.1\times 10^{5} $\\
 Mass of each test body set & \centering{$m$}  &\ $  19.4 $ g\\
 Lever-arm & \centering{$r$}  &\ $  1.9 $ cm\\
 Moment of inertia & \centering{$I$}  &\ $  3.78 \times 10^{-5}$ kg $\text{m}^2$\\
 Resonant frequency & \centering{$\omega_0$}  &\ $  2\pi \times 0.69$ mHz\\  
 Turntable frequency & \centering{$\omega_{TT}$}  &\ $  2\pi \times 0.46$ mHz\\
 \hline
\end{tabular}
 \caption{Relevant parameters of the torsion balance. The lever-arm $r$ is the radial distance from the pendulum's fiber axis to the center of mass of each set of test bodies.}\label{appTable}
\end{table}

\section{Analysis} 

The experiment recorded data from July 7, 2024 to December 25, 2025. We rotated the pendulum $180^\circ$ with respect to the vacuum chamber on July 29, 2024, February 15, 2025, May 1, 2025, and September 29, 2025 to reduce systematic effects that may be associated with the pendulum orientation relative to the vacuum chamber (such as tilt and electrostatics). Rotating the pendulum changes the sign of the science signal while maintaining most systematics; these pendulum reorientations were accounted for in the analysis.  

The duty cycle of the apparatus was severely restricted due to nearby construction activity and an unexpected hardware failure. Although the data spans 537 days, only 244 days (45\%) were free of external disturbances and included in the data set. The measured torsion balance angle was calibrated with a programmed turntable speed change (see Supplementary Material). The angle was converted to differential acceleration by inverting the harmonic oscillator response of the pendulum with a time-domain filter and scaling by the relevant fixed parameters. An additional low-pass filter (1~mHz third-order Butterworth) was applied to minimize the effect of high frequency noise. We corrected for the response of this filter at the turntable frequency (0.4\% attenuation, $\sim54^\circ$ phase). The differential acceleration time series was split into two turntable-rotation long sets. Each set was fit to a linear combination of sinusoidal functions with frequencies of harmonics of the turntable angle ($\omega_{TT}$, $2\omega_{TT}$, $3\omega_{TT}$, etc.) and harmonics of the torsional resonance ($\omega_0$, $2\omega_0$, $3\omega_0$, etc.). The full fit function is shown in the Supplementary Material.

\begin{figure}[!h]
\centering \includegraphics[width=0.5\textwidth]{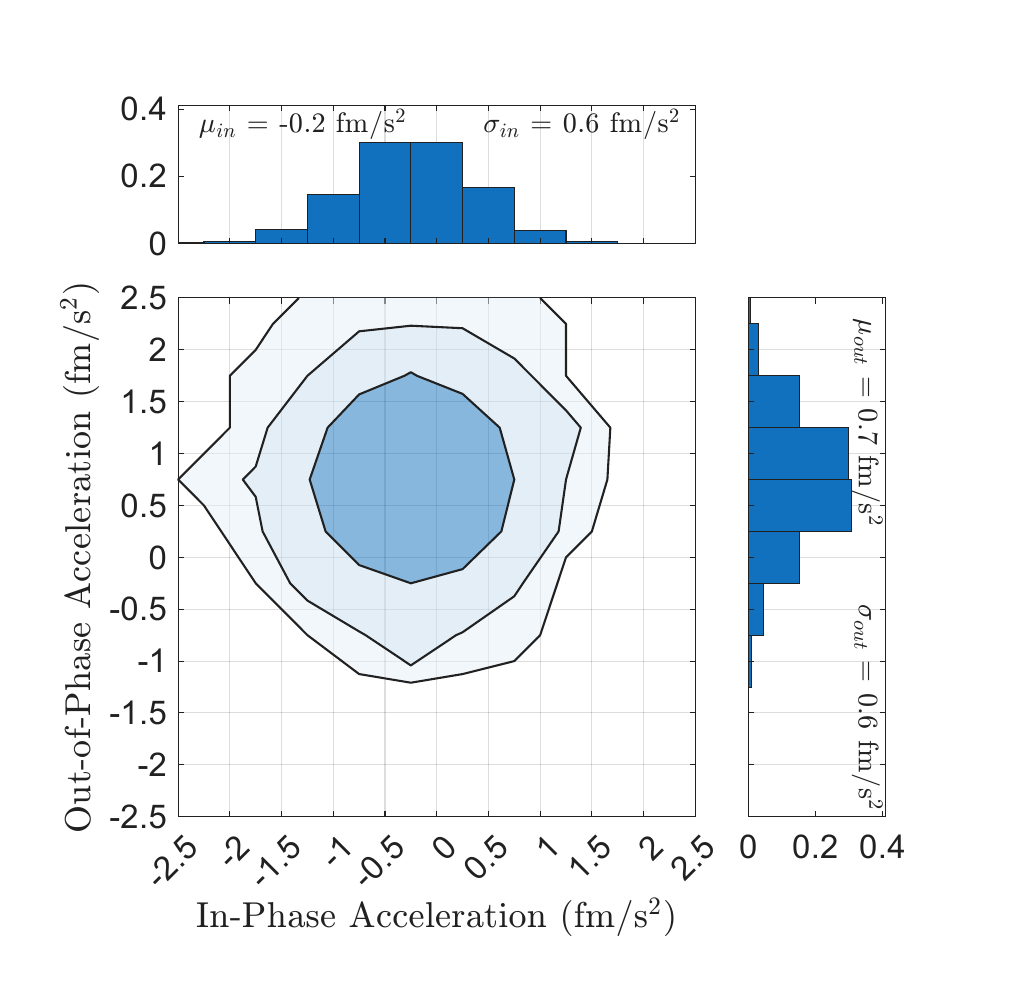}
\caption{Distributions of the accelerations towards the galactic center (in-phase) and the orthogonal direction (out-of-phase). The top and right panels show the histograms for the in-phase and out-of-phase accelerations, respectively. These distributions were calculated by Monte Carlo sampling of the $\omega_{TT}$-acceleration measurements repeated 1000 times.}
\label{fits} 
\end{figure}

\pagebreak

The instrument experienced occasional transients due to nearby construction activity and environmental disturbances as well as seismic activity. We discarded data sets that fell under two criteria: misfit-squared exceeding the 99th percentile of the corresponding $\chi^2$-distribution, and a once per turntable-revolution (${\omega_{TT}}$) acceleration component outside of the 95th percentile of the component's distribution. We subtracted the mean ${\omega_{TT}}$-acceleration amplitude of each data set leaving only the time varying components. Figure \ref{time} shows the surviving once per turntable revolution acceleration amplitudes. 

The $\omega_{TT}$-acceleration was then fit to a set of galactic basis functions to yield the differential acceleration towards the galactic dark matter. The galactic functions consisted of the location of the center of the Milky Way (RA: $18^\text{h}$ Dec: $-29^\circ$) projected into the horizontal plane of the lab and an orthogonally oriented function. These functions modulate once per sidereal day due to the rotation of the Earth and the location of our lab at $\sim48^\circ$ latitude. The location of the galactic center and the necessary coordinate transformations were computed using the {\tt Astropy} \cite{astropy:2022} libraries.

We deployed a Monte Carlo resampling technique to assess the uncertainties on the best-fit galactic acceleration amplitudes. This consisted of refitting a randomly sampled set of the $\omega_{TT}$-accelerations to the galactic basis functions. The random sets contained equal number of data points as the initial set. This was repeated 1000 times to yield the distributions shown in Figure \ref{fits}.

Additionally, we conducted an independent cut-based analysis which verified the results. The cut-based analysis is detailed in the Supplemental Material.

\section{Results} 

Measuring the out-of-phase component verifies the sensitivity of the experiment as the signal of interest would only appear as a non-zero in-phase component. We take the mean and corresponding uncertainty calculated from the scatter of the measurements to give a final measure of the differential acceleration:
\begin{equation}
    \Delta a _{\text{DM, Be-Al}}= -0.2 \pm 0.6 \text{ fm/s}^2\quad (1\sigma) 
\end{equation} 
A positive $\Delta a$ would correspond to beryllium attracting more towards dark matter.
The average acceleration for the $0^\circ$ and $180^\circ$ orientations are individually consistent with zero. 

The gravitational acceleration on Earth towards the galactic dark matter is $\geq$50 $\text{pm/s}^2$ \cite{PhysRevLett.70.119}. The acceleration towards ordinary visible matter is approximately four times larger, but previous measurements comparing the accelerations of beryllium and aluminum towards the Earth have shown no novel interactions to the $~\sim~10^{-13}$ level. Therefore, any acceleration towards the center of the galaxy that distinguishes between beryllium and aluminum must be entirely due to dark matter and must be weaker than the gravitational acceleration from dark matter by a factor of:
\begin{equation}
    \eta_{\text{DM, Be-Al}} \leq 2.4 \times 10^{-5}\quad (95\%\text{-confidence}). \label{eta}
\end{equation}
These constraints are a factor of four improvement over previous measurements \cite{wagner2012torsion}.

\section{Discussion} Our observations place limits on the strength of any galactic-range interaction between dark matter and normal matter that does not couple identically to aluminum and beryllium \cite{PhysRevLett.70.119}. Other experiments have made different material choices and thus restrict these finely-tuned scenarios as well \cite{wagner2012torsion, PhysRevD.100.084030}.


There are no underlying symmetries in the Standard Model that require conservation of baryon number (B) and lepton number (L).  Yet, global violation of these conservations have not been observed. This has led to speculation that there may exist a B- or L-dependent fifth force \cite{Broussard_2025, Dev_2024, perez2022baryon, 10.3389/fphy.2018.00040}. Such fifth forces have been stringently tested for normal matter \cite{PhysRevLett.129.121102}. Our results extend these tests to include dark matter-normal matter interactions.

Grand Unified Theories \cite{DEBOER1994201} attempt to unify the electromagnetic, weak, and strong forces of nature into a single force. In many of these theories, B and L numbers are not conserved; instead the difference (B-L) is the conserved quantity \cite{BUCHMULLER1991395, PhysRevD.80.055030}. This conservation has many possible consequences for dark matter including that dark matter may have B-L charge \cite{Rodejohann_2015, PhysRevD.91.115017}. Our experiment is uniquely designed to search for interactions to this B-L charged dark matter; the test body materials were chosen to maximize their difference in B-L charge. There is ambiguity in what B-L charge dark matter could have. Regardless, our results restrict the strength of such an interaction.

The equivalence principle is a foundational aspect of our geometric understanding of gravity (i.e. General Relativity). The principle states that gravity is independent of the composition of objects. This has been tested extensively for a variety of materials and distance scales~\cite{PhysRevLett.129.121102, wagner2012torsion}.  However, these tests use normal matter as the source mass, making them blind to any violation associated with dark matter. Our results directly test the equivalence principle towards dark matter at galactic scales and set limits on the E\"otv\"os parameter for beryllium and aluminum towards dark matter ($\eta_\text{DM, Be-Al}$ in Equation \ref{eta}).

\section{Conclusion} 

We operated a rotating torsion balance to search for a galactic-range interaction between normal matter and dark matter. We observe no evidence for such an interaction and set strict limits on its possible strength. Although this constrains a variety of theoretical possibilities for dark matter, our observation is model independent. Any proposed dark matter interaction that does not couple identically to beryllium and aluminum is constrained by this observation.

There are very few laboratory experiments that directly probe dark matter without assuming a specific dark matter theory. Our experiment achieved this on a galactic scale while being located on Earth. This observation furthers our understanding of not just dark matter but also the nature of gravity and fundamental symmetries of the universe.

\section{Data Availability} 

Raw data will be shared upon request. The fit amplitudes and analysis code can be found at \url{https://github.com/EotWash/EP-Analysis}.

\section{Acknowledgements} 

This work was supported by funding from the NSF under Awards PHY-1607385, PHY-1607391, PHY-1912380, and PHY-1912514.

\bibliography{main.bib}

@article{shaw2022torsion,
  title={Torsion-balance search for ultralow-mass bosonic dark matter},
  author={Shaw, EA and Ross, MP and Hagedorn, CA and Adelberger, EG and Gundlach, JH},
  journal={Physical Review D},
  volume={105},
  number={4},
  pages={042007},
  year={2022},
  publisher={APS}
}

@article{thermal,
  title = {Thermal noise in mechanical experiments},
  author = {Saulson, Peter R.},
  journal = {Phys. Rev. D},
  volume = {42},
  issue = {8},
  pages = {2437--2445},
  numpages = {0},
  year = {1990},
  month = {Oct},
  publisher = {American Physical Society},
  doi = {10.1103/PhysRevD.42.2437},
  url = {https://link.aps.org/doi/10.1103/PhysRevD.42.2437}
}

@book{shaw2023equivalence,
  title={Equivalence Principle Tests and Direct Searches for Ultra-Light Dark Matter with Fused-Silica Torsion Fibers},
  author={Shaw, Erik A},
  year={2023},
  publisher={University of Washington}
}

@article{wagner2012torsion,
  title={Torsion-balance tests of the weak equivalence principle},
  author={Wagner, Todd A and Schlamminger, Stephan and Gundlach, Jens H and Adelberger, Eric G},
  journal={Classical and Quantum Gravity},
  volume={29},
  number={18},
  pages={184002},
  year={2012},
  publisher={IOP Publishing}
}

@ARTICLE{astropy:2022,
       author = {{Astropy Collaboration} and {Astropy Project Contributors}},
        title = "{The Astropy Project: Sustaining and Growing a Community-oriented Open-source Project and the Latest Major Release (v5.0) of the Core Package}",
      journal = {\apj},
         year = 2022,
        month = aug,
       volume = {935},
       number = {2},
          eid = {167},
        pages = {167},
          doi = {10.3847/1538-4357/ac7c74},
archivePrefix = {arXiv},
       eprint = {2206.14220},
 primaryClass = {astro-ph.IM},
       adsurl = {https://ui.adsabs.harvard.edu/abs/2022ApJ...935..167A}
}

@book{bertone2010particle,
  title={Particle dark matter: observations, models and searches},
  author={Bertone, Gianfranco},
  year={2010},
  publisher={Cambridge University Press}
}

@article{arcadi2018waning,
  title={The waning of the WIMP? A review of models, searches, and constraints},
  author={Arcadi, Giorgio and Dutra, Ma{\'\i}ra and Ghosh, Pradipta and Lindner, Manfred and Mambrini, Yann and Pierre, Mathias and Profumo, Stefano and Queiroz, Farinaldo S},
  journal={The European Physical Journal C},
  volume={78},
  pages={1--57},
  year={2018},
  publisher={Springer}
}

@article{PhysRevLett.70.119,
  title = {Experimental limits on any long range nongravitational interaction between dark matter and ordinary matter},
  author = {Stubbs, Christopher W.},
  journal = {Phys. Rev. Lett.},
  volume = {70},
  issue = {2},
  pages = {119--122},
  numpages = {0},
  year = {1993},
  month = {Jan},
  publisher = {American Physical Society},
  doi = {10.1103/PhysRevLett.70.119},
  url = {https://link.aps.org/doi/10.1103/PhysRevLett.70.119}
}

@article{sterile,
title = {Sterile neutrino Dark Matter},
journal = {Progress in Particle and Nuclear Physics},
volume = {104},
pages = {1-45},
year = {2019},
issn = {0146-6410},
doi = {https://doi.org/10.1016/j.ppnp.2018.07.004},
url = {https://www.sciencedirect.com/science/article/pii/S0146641018300711},
author = {A. Boyarsky and M. Drewes and T. Lasserre and S. Mertens and O. Ruchayskiy},
}

@ARTICLE{PBH,
  
AUTHOR={Villanueva-Domingo, Pablo  and Mena, Olga  and Palomares-Ruiz, Sergio },
         
TITLE={A Brief Review on Primordial Black Holes as Dark Matter},
        
JOURNAL={Frontiers in Astronomy and Space Sciences},
        
VOLUME={Volume 8 - 2021},

YEAR={2021},

URL={https://www.frontiersin.org/journals/astronomy-and-space-sciences/articles/10.3389/fspas.2021.681084},

DOI={10.3389/fspas.2021.681084},

ISSN={2296-987X},
}

@article{ planck,
	author = {Planck Collaboration},
	title = {{Planck 2018 results - I. Overview and the cosmological legacy of Planck}},
	DOI= "10.1051/0004-6361/201833880",
	url= "https://doi.org/10.1051/0004-6361/201833880",
	journal = {A\&A},
	year = 2020,
	volume = 641,
	pages = "A1",
}

@article{des,
doi = {10.3847/2041-8213/ad6f9f},
url = {https://dx.doi.org/10.3847/2041-8213/ad6f9f},
year = {2024},
month = {oct},
publisher = {The American Astronomical Society},
volume = {973},
number = {1},
pages = {L14},
author = {DES Collaboration},
title = {{The Dark Energy Survey: Cosmology Results with 1500 New High-redshift Type Ia Supernovae Using the Full 5 yr Data Set}},
journal = {The Astrophysical Journal Letters},
}

@article{desi,
      title={{DESI DR2 Results II: Measurements of Baryon Acoustic Oscillations and Cosmological Constraints}}, 
      author={DESI Collaboration},
      journal = {arxiv},
      year={2025},
      eprint={2503.14738},
      archivePrefix={arXiv},
      primaryClass={astro-ph.CO},
      url={https://arxiv.org/abs/2503.14738} 
}

@article{rotcurve,
  title={The distribution of dark matter in galaxies},
  author={Salucci, Paolo},
  journal={The Astronomy and Astrophysics Review},
  volume={27},
  number={1},
  pages={2},
  year={2019},
  publisher={Springer}
}

@unpublished{milkyway,
    author = {ESA/Gaia/DPAC, Stefan Payne-Wardenaar},
    year = {2025},
    title = {{The best Milky Way map, by Gaia (edge-on)}},
    url = {https://www.esa.int/ESA_Multimedia/Images/2025/01/The_best_Milky_Way_map_by_Gaia_edge-on}
}

@article{Broussard_2025,
doi = {10.1088/1361-6471/adf081},
url = {https://dx.doi.org/10.1088/1361-6471/adf081},
year = {2025},
month = {aug},
publisher = {IOP Publishing},
volume = {52},
number = {8},
pages = {083001},
author = {Broussard, Leah J and others},
title = {{Baryon number violation: from nuclear matrix elements to BSM physics}},
journal = {Journal of Physics G: Nuclear and Particle Physics}
}

@article{Dev_2024,
doi = {10.1088/1361-6471/ad1658},
url = {https://dx.doi.org/10.1088/1361-6471/ad1658},
year = {2024},
month = {jan},
publisher = {IOP Publishing},
volume = {51},
number = {3},
pages = {033001},
author = {Dev, P S B and others},
title = {Searches for baryon number violation in neutrino experiments: a white paper},
journal = {Journal of Physics G: Nuclear and Particle Physics}
}

@article{perez2022baryon,
  title={On baryon and lepton number violation},
  author={Perez, Pavel Fileviez and Pocar, Andrea and Babu, KS and Broussard, Leah J and Cirigliano, Vincenzo and Gardner, Susan and Heeck, Julian and Kearns, Ed and Long, Andrew J and Raby, Stuart and others},
  journal={arXiv preprint arXiv:2208.00010},
  year={2022}
}

@ARTICLE{10.3389/fphy.2018.00040,
  
AUTHOR={Cai, Yi  and Han, Tao  and Li, Tong  and Ruiz, Richard },
         
TITLE={Lepton Number Violation: Seesaw Models and Their Collider Tests},
        
JOURNAL={Frontiers in Physics},
        
VOLUME={Volume 6 - 2018},

YEAR={2018},

URL={https://www.frontiersin.org/journals/physics/articles/10.3389/fphy.2018.00040},

DOI={10.3389/fphy.2018.00040},

ISSN={2296-424X}}

@article{PhysRevLett.129.121102,
  title = {{$MICROSCOPE$ Mission: Final Results of the Test of the Equivalence Principle}},
  author = {Touboul, Pierre and others},
  collaboration = {MICROSCOPE Collaboration},
  journal = {Phys. Rev. Lett.},
  volume = {129},
  issue = {12},
  pages = {121102},
  numpages = {8},
  year = {2022},
  month = {Sep},
  publisher = {American Physical Society},
  doi = {10.1103/PhysRevLett.129.121102},
  url = {https://link.aps.org/doi/10.1103/PhysRevLett.129.121102}
}

@article{DEBOER1994201,
title = {Grand unified theories and supersymmetry in particle physics and cosmology},
journal = {Progress in Particle and Nuclear Physics},
volume = {33},
pages = {201-301},
year = {1994},
issn = {0146-6410},
doi = {https://doi.org/10.1016/0146-6410(94)90045-0},
url = {https://www.sciencedirect.com/science/article/pii/0146641094900450},
author = {W {De Boer}}
}

@article{Rodejohann_2015,
doi = {10.1088/1475-7516/2015/12/032},
url = {https://dx.doi.org/10.1088/1475-7516/2015/12/032},
year = {2015},
month = {dec},
publisher = {},
volume = {2015},
number = {12},
pages = {032},
author = {Rodejohann, Werner and Yaguna, Carlos E.},
title = {{Scalar dark matter in the B-L model}},
journal = {Journal of Cosmology and Astroparticle Physics}
}

@article{PhysRevD.91.115017,
  title = {{Accidental dark matter: Case in the scale invariant local $B-L$ model}},
  author = {Guo, Jun and Kang, Zhaofeng and Ko, P. and Orikasa, Yuta},
  journal = {Phys. Rev. D},
  volume = {91},
  issue = {11},
  pages = {115017},
  numpages = {15},
  year = {2015},
  month = {Jun},
  publisher = {American Physical Society},
  doi = {10.1103/PhysRevD.91.115017},
  url = {https://link.aps.org/doi/10.1103/PhysRevD.91.115017}
}

@article{BUCHMULLER1991395,
title = {{Neutrino masses, neutral vector bosons and the scale of B-L breaking}},
journal = {Physics Letters B},
volume = {267},
number = {3},
pages = {395-399},
year = {1991},
issn = {0370-2693},
doi = {https://doi.org/10.1016/0370-2693(91)90952-M},
url = {https://www.sciencedirect.com/science/article/pii/037026939190952M},
author = {W. Buchm\"uller and C. Greub and P. Minkowski}
}

@article{PhysRevD.80.055030,
  title = {{Phenomenology of the minimal $B\ensuremath{-}L$ extension of the standard model: ${Z}^{\ensuremath{'}}$ and neutrinos}},
  author = {Basso, Lorenzo and Belyaev, Alexander and Moretti, Stefano and Shepherd-Themistocleous, Claire H.},
  journal = {Phys. Rev. D},
  volume = {80},
  issue = {5},
  pages = {055030},
  numpages = {14},
  year = {2009},
  month = {Sep},
  publisher = {American Physical Society},
  doi = {10.1103/PhysRevD.80.055030},
  url = {https://link.aps.org/doi/10.1103/PhysRevD.80.055030}
}

@article{HUNT2025101721,
title = {Milky Way dynamics in light of Gaia},
journal = {New Astronomy Reviews},
volume = {100},
pages = {101721},
year = {2025},
issn = {1387-6473},
doi = {https://doi.org/10.1016/j.newar.2024.101721},
url = {https://www.sciencedirect.com/science/article/pii/S1387647324000289},
author = {Jason A.S. Hunt and Eugene Vasiliev}
}

@article{ADELBERGER2009102,
title = {Torsion balance experiments: A low-energy frontier of particle physics},
journal = {Progress in Particle and Nuclear Physics},
volume = {62},
number = {1},
pages = {102-134},
year = {2009},
issn = {0146-6410},
doi = {https://doi.org/10.1016/j.ppnp.2008.08.002},
url = {https://www.sciencedirect.com/science/article/pii/S0146641008000720},
author = {E.G. Adelberger and J.H. Gundlach and B.R. Heckel and S. Hoedl and S. Schlamminger}
}

@article{BOZORGNIA2024,
title = {Dark matter candidates and searches},
journal = {Canadian Journal of Physics},
year = {2024},
issn = {0008-4204},
doi = {https://doi.org/10.1139/cjp-2024-0128},
url = {https://www.sciencedirect.com/science/article/pii/S0008420424000692},
author = {Nassim Bozorgnia and Joseph Bramante and James M. Cline and David Curtin and David McKeen and David E. Morrissey and Adam Ritz and Simon Viel and Aaron C. Vincent and Yue Zhang}
}

@article{PhysRevD.100.084030,
  title = {Constraints on fifth forces through perihelion precession of planets},
  author = {Sun, Bing and Cao, Zhoujian and Shao, Lijing},
  journal = {Phys. Rev. D},
  volume = {100},
  issue = {8},
  pages = {084030},
  numpages = {7},
  year = {2019},
  month = {Oct},
  publisher = {American Physical Society},
  doi = {10.1103/PhysRevD.100.084030},
  url = {https://link.aps.org/doi/10.1103/PhysRevD.100.084030}
}

\section{Supplementary Material}

\textit{Fit Functions} - The observed differential acceleration was fit to the following linear combination of sinusoidal functions with frequencies of harmonics of the turntable angle ($\omega_{TT}$, $2\omega_{TT}$, $3\omega_{TT}$, etc.) and harmonics of the torsional resonance ($\omega_0$, $2\omega_0$, $3\omega_0$, etc.):
\begin{align}\nonumber
    \Delta a & =  a_{TT} \cos(\omega_{TT} t) + b_{TT} \sin(\omega_{TT} t) \\ \nonumber
    & + a_{2TT} \cos(2\omega_{TT} t) + b_{2TT} \sin(2\omega_{TT} t)\\ \nonumber
    & + a_{3TT} \cos(3\omega_{TT} t) + b_{3TT} \sin(3\omega_{TT} t)\\ \nonumber
    & + a_{4TT} \cos(4\omega_{TT} t) + b_{4TT} \sin(4\omega_{TT} t)\\ \nonumber
    & + a_{5TT} \cos(5\omega_{TT} t) + b_{5TT} \sin(5\omega_{TT} t)\\ \nonumber
    & +  a_{1} \cos(\omega_0 t) + b_{1} \sin(\omega_0 t)\\ \nonumber
    & +  a_{2} \cos(2\omega_0 t) + b_{2} \sin(2\omega_0 t)\\ \nonumber
    & +  a_{3} \cos(3\omega_0 t) + b_{3} \sin(3\omega_0 t)\\ \nonumber
    & +  a_{4} \cos(4\omega_0 t) + b_{4} \sin(4\omega_0 t)\\ \nonumber
    & +  a_{5} \cos(5\omega_0 t) + b_{5} \sin(5\omega_0 t)\\ \nonumber
    & + K  \nonumber
\end{align}
where $a_i$ and $b_i$ are the harmonic fit parameters, $t$ is time, $K$ is the offset fit parameter.

\begin{figure}[!h]
\centering \includegraphics[width=0.45\textwidth]{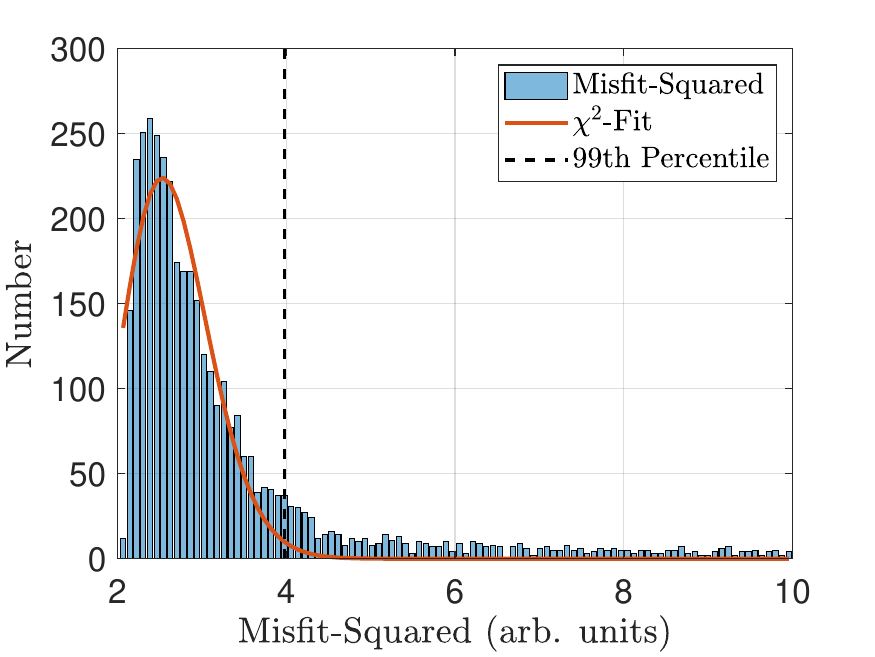}
\caption{Misfit-squared, fit to a $\chi^2$ distribution, and 99th-percentile cutoff.}
\label{chi} 
\end{figure}

The once per turn table revolution amplitudes, $a_{TT}$ and $b_{TT}$, where combined into a single complex number $a_{TT}+i b_{TT}$ whose amplitude is shown in Figure \ref{time}. These were then fit to the galactic basis functions, $z_{\text{in}}$ and $z_{\text{out}}$ via:
\begin{align}
    a_{TT}+i b_{TT} = (\alpha +i\beta)\bigg( z_{\text{in}}(t)+i z_{\text{out}}(t)\bigg)
\end{align}
to give the in-phase and out-of-phase amplitudes $\alpha$ and $\beta$, respectively. The use of complex numbers allows the phases of the various signals to be readily accounted for.

\textit{$\chi^2$-Cuts} - We discarded data that was contaminated by external disturbances. To do this, we fit the misfit-squared of the data sets to a $\chi^2$-distribution, shown in Figure \ref{chi}. The 99th-percentile of this distribution was then used as a cutoff. Any data set whose misfit-squared was greater than this was discarded.

\begin{figure}[!h]
\centering \includegraphics[width=0.45\textwidth]{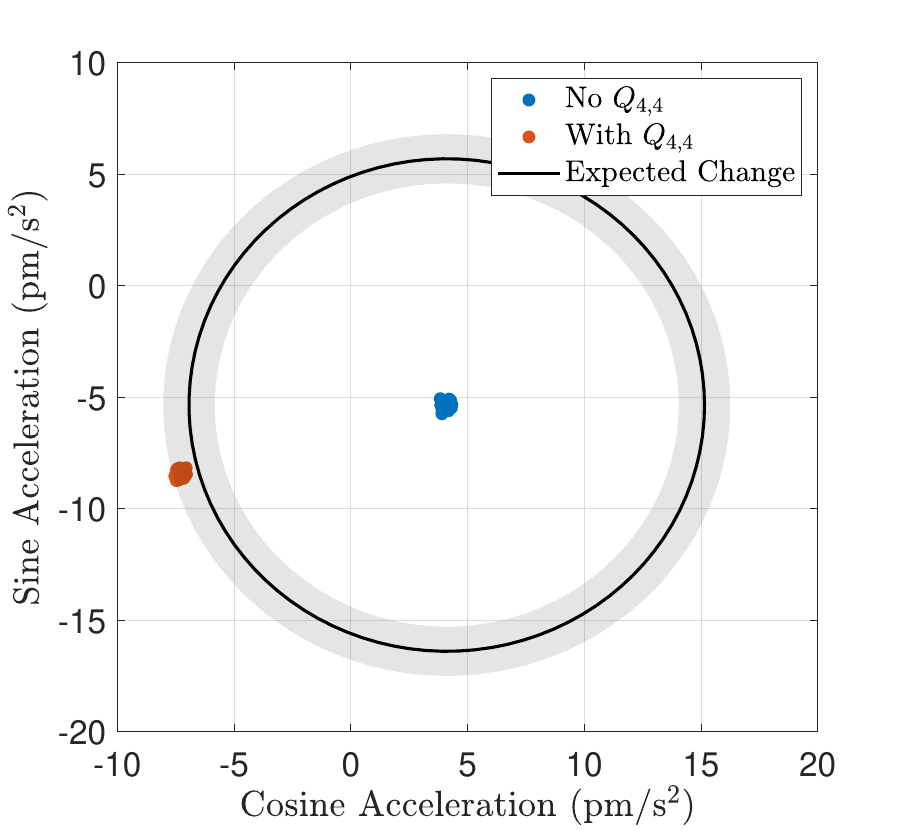}
\caption{Extracted $4\omega_{TT}$ accelerations with and without the gravitational $Q_{4,4}$ masses along with the expected change in amplitude.}
\label{Q44} 
\end{figure}

\textit{Sensitivity Verification} - The torsion pendulum had a significant $q_{44}$ moment due to it's four-fold geometry. We utilized this moment to verify the acceleration sensitivity of the apparatus. A set of four 1.35 kg brass masses were placed in-plane with the pendulum at a radius of 28.6 cm in a four fold symmetric pattern. This induced a known external $Q_{44}$. The coupling between the induced $Q_{44}$ and the pendulum's $q_{44}$ produced an acceleration at $4\omega_{TT}$.

The measurement of the $4\omega_{TT}$ acceleration amplitude with and without the injected $Q_{44}$ is shown in Figure \ref{Q44} along with the expected change in  $4\omega_{TT}$ acceleration. The environment had a small $Q_{44}$ without the injection. Without tracking the phase the expected signal must be on a ring of fixed amplitude around the environmental $4\omega_{TT}$ acceleration.

 \begin{figure}[!h]
\centering \includegraphics[width=0.45\textwidth]{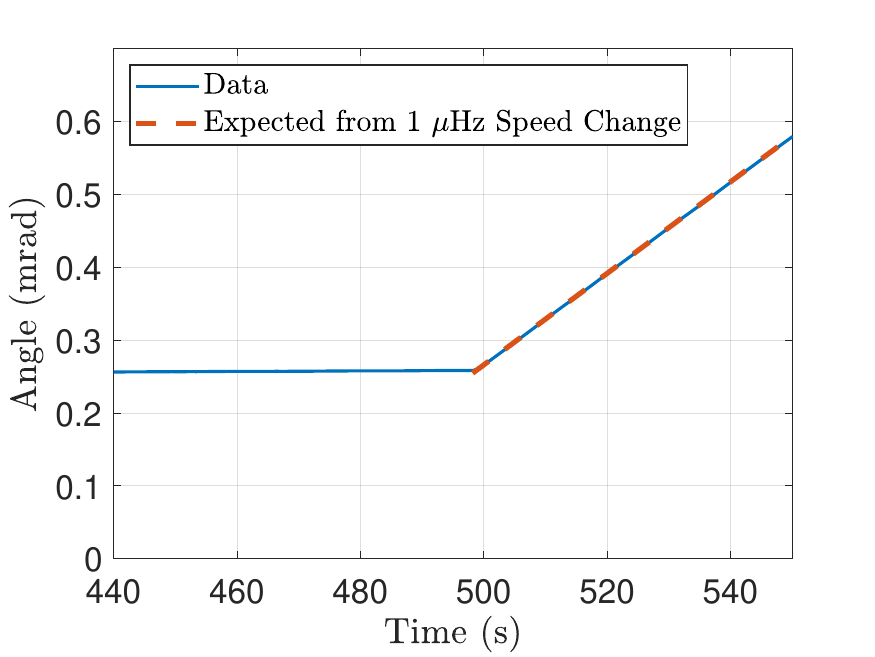}
\caption{Autocollimator angle readings during a 1 {\textmu}Hz turntable speed change. This speed change allows the angle calibration to be cross-checked with the turntable.}
\label{cal} 
\end{figure}

\textit{Inertial Calibration} - The air-bearing turntable's rotation rate was fed back to a 36000 stripes per revolution angle encoder with two readheads.  The autocollimator was rigidly attached to the vacuum chamber, which was rotated by the turntable.  We utilized this to calibrate the angular readout relative to the turntable speed. 

\begin{figure}[!h]
\centering \includegraphics[width=0.45\textwidth]{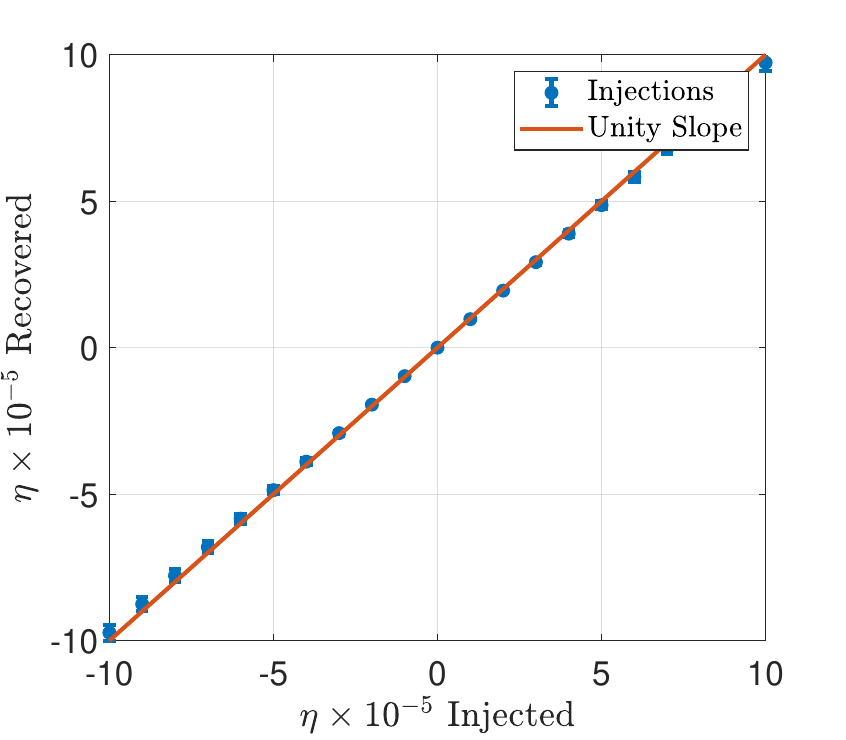}
\caption{False dark matter signal injection and extraction. }
\label{inj} 
\end{figure}

When the turntable is subjected to a sudden a step change in speed, the pendulum stays practically inertial while the autocollimator begins to rotate. This appears as an approximately linear angle change whose slope depends on the step change amplitude as shown in Figure \ref{cal}

\textit{Signal Injection} - We conducted a set of artificial signal injections through our analysis to ensure accurate extraction of a non-zero $\eta$. These were done from $\eta = -10^{-4}$ to $10^{-4}$ is $10^{-5}$ steps. The results of these injections are shown in Figure \ref{inj}.

This shows accurate extraction of the signal across a wide selection of $\eta$ values. The slight non-linear behavior at large injection amplitudes is due to the statistical cuts removing the peaks of the injected signal. This behavior is eliminated by increasing the misfit threshold.

\textit{Tilt Coupling} - The tilt of the apparatus relative to local vertical can introduce false once per revolution signals. We employed a control system to minimize this effect. However, residual tilts may affect our results.

We conducted a series of tilt tests to place bounds on this effect. The tilt was injected by oscillating the temperature of one of the thermal expansion support feet at 0.57 mHz. The resultant tilt and apparent differential acceleration were then compared to get a coupling from tilt to acceleration. 

This coupling can differ from tilt direction and pendulum mirror. We measured the coupling of each mirror used in the science data ($0^\circ$ and $180^\circ$) for each tilt direction (A and B)

\begin{align} \nonumber
    K_{0^\circ,A} &= 6.3 \pm 0.3 \text{ nm/s}^2/\text{rad}\\ \nonumber
    K_{0^\circ,B} &= 25.2 \pm 0.5\text{ nm/s}^2/\text{rad}\\ \nonumber
    K_{180^\circ,A} &= 28.2 \pm 1.1 \text{ nm/s}^2/\text{rad}\\ \nonumber
    K_{180^\circ,B} &= 24.4 \pm 0.5 \text{ nm/s}^2/\text{rad}\nonumber 
\end{align}
The apparatus tilts $\sim$ 4.7 nrad per day. Using this, we place an upper limit of the possible tilt induced differential acceleration of $\Delta a_{\text{tilt}} \leq 0.3 \text{ fm/s}^2\ (\eta_\text{tilt} \leq 0.6 \times 10^{-5})$.

\textit{Magnetic Coupling} - The influence of external magnetic fields is significantly reduced by the magnetic shielding. We conducted a set external of magnetic injections to place limits on the residual influence. This consisted of placing large coils around the apparatus through which an oscillating current was driven at 0.57 mHz. A large external field of 70 {\textmu}G showed no influence on the pendulum. With this non-observation, we can set an upper limit on the magnetic coupling of $0.14~\text{nm/s}^2/\text{G}$ and a maximum magnetic systematic of $\Delta a_\text{mag} \leq 1.1 \times 10^{-3} \text{ fm/s}^2\ (\eta_\text{mag} \leq 0.006 \times 10^{-5})$ .

\begin{figure}[!h]
\centering \includegraphics[width=0.5\textwidth]{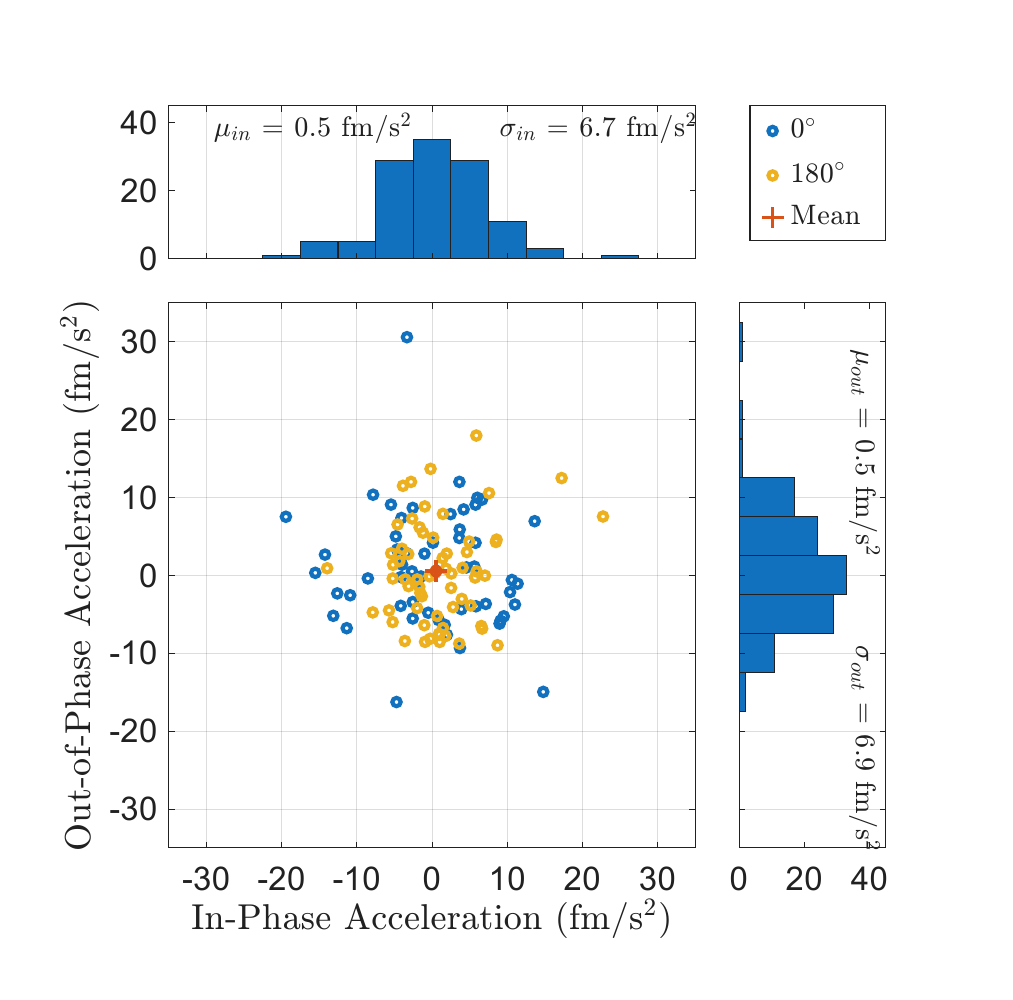}
\caption{Scatter plot of the accelerations towards the galactic center (in-phase) and the orthogonal
direction (out-of-phase) for the cut-based analysis. }
\label{galCut} 
\end{figure}

\textit{Gravity Gradient Coupling} - The design of the pendulum minimizes its gravitational moments. We measured the residual $q_{2,1}$ moment by rotating the exterior $Q_{2,1}$ compensating masses by 180$^\circ$. This yielded a $q_{2,1} = 0.03 \text{ g/cm}^2$.

\textit{Cut-based Analysis} - We performed a secondary cut-based analysis to verify the analysis techniques described in the main body of the paper. This consisted of extracting the $\omega_{TT}$-acceleration amplitudes in two-day long segments. Segments that had less than a quarter of a sidereal day of data were discarded. Each segment was fit to the matching segment of the galactic basis functions, as shown in Figure 10. This fitting yielded 56 points for the 0$^\circ$ pendulum orientation and 66 points for 180$^\circ$ for a total of 122 independent measurements.

The series of measured differential accelerations toward the galactic center is shown in Figure \ref{galCut}. We take the mean and corresponding uncertainty calculated from the scatter of the measurements to give a final measure of the differential acceleration:
\begin{equation}
    \Delta a _\text{DM, Be-Al}= 0.5 \pm 0.6 \text{ fm/s}^2\quad (1\sigma) 
\end{equation} 

\begin{widetext}

\begin{figure}[!h]
\centering \includegraphics[width=\textwidth]{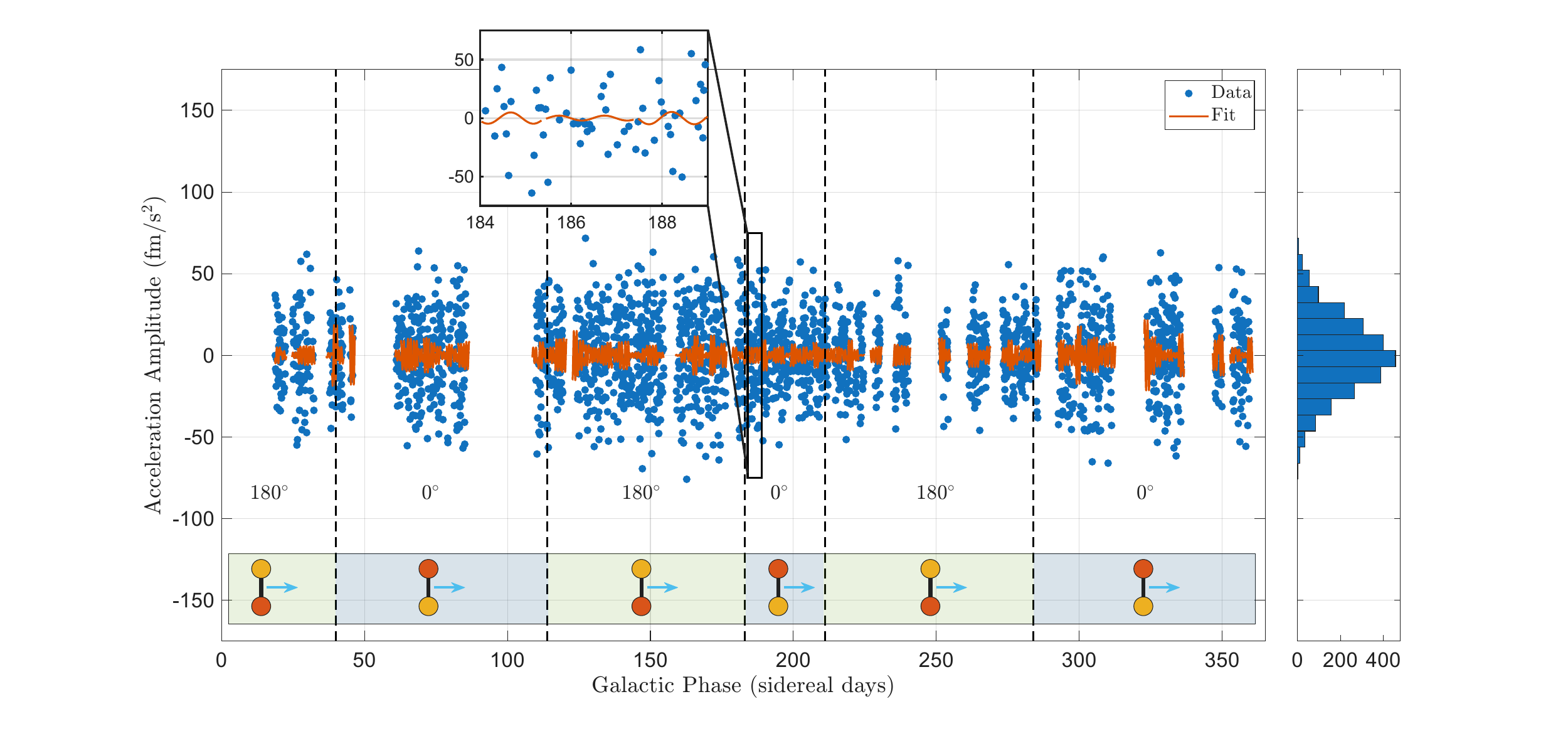}
\caption{Once per turntable revolution differential acceleration amplitudes along with the corresponding fits to the galactic basis functions for the cut-based analysis.}
\label{timCut} 
\end{figure}

\end{widetext}

\end{document}